\documentstyle[prc,aps,epsfig]{revtex}

\begin{document}
\title{High-precision measurement of the half-life of $^{62}$Ga}
\author
 {B. Blank$^{1,2}$, G. Savard$^{1}$, J. D\"oring$^{3}$, A. Blazhev$^{3,4}$, 
  G. Canchel$^{2}$, M. Chartier$^{5}$, D. Henderson$^{1}$, Z. Janas$^{6}$, 
  R. Kirchner$^{3}$, I. Mukha$^{3}${\footnotemark}, E. Roeckl$^{3}$, K. Schmidt$^{7}$,
  J. \.{Z}ylicz$^{6}$
  }

\address{
$^{1}$Physics Division, Argonne National Laboratory, Argonne, IL 60439, USA \\
$^{2}$Centre d'Etudes Nucl\'eaires de Bordeaux-Gradignan, Le Haut-Vigneau,
      B.P. 120, F-33175 Gradignan, France \\
$^{3}$GSI Darmstadt, Planckstrasse 1, D-64291 Darmstadt, Germany\\
$^{4}$University of Sofia, BG-1164 Sofia, Bulgaria \\
$^{5}$Oliver Lodge Laboratory, University of Liverpool, Liverpool L69 7ZE, UK\\
$^{6}$Institute of Experimental Physics, University of Warsaw, 
      PL-00-681 Warsaw, Poland \\
$^{7}$Department of Physics \& Astronomy, University of Edinburgh,
      Edinburgh EH9 3JZ, UK \\
}

\maketitle

\begin{abstract}
The $\beta$-decay half-life of $^{62}$Ga has been studied with high precision
using on-line mass separated samples.
The decay of $^{62}$Ga which is dominated by a $0^+$ to $0^+$ transition to the
ground state of $^{62}$Zn yields a half-life of  $T_{1/2}$ = 116.19(4) ms. 
This result is more precise than any previous measurement by about a factor
of four or more. The present value is in agreement with older literature 
values, but slightly disagrees with a recent measurement. We determine
an error weighted average value of all experimental half-lives of 116.18(4)~ms.

\end{abstract}

\vspace*{0.5cm}
\begin{center}

{\bf PPACs: 23.40.-s, 21.10.Tg, 27.50.+e}

\end{center}

\renewcommand{\thefootnote}{\fnsymbol{footnote}}

\footnotetext[1]{Present address: Instituut voor Kern-en Stralingsfysica,
Katholic University of Leuven, B-3001 Leuven, Belgium}

\section{Introduction}

The Fermi coupling constant $G_v$ is most precisely determined from the nuclear 
$\beta$ decay of nine isotopes ranging from $^{10}$C to 
$^{54}$Co~\cite{towner02,hardy90}.
The $0^+$ to $0^+$ Fermi decays of these isotopes allow to determine
a corrected $Ft$ value from the ft values of the individual isotopes 
and thus the coupling constant $G_v$ by means of the relation~\cite{towner02}
\begin{equation}
Ft = ft(1 + \delta^{'}_R)(1 + \delta_{NS} - \delta_C) = \frac{k}{2 G_v(1+\Delta_R^v)}.
\end{equation}

$f$ is the statistical rate function which strongly depends on the decay 
Q value for $\beta$ decay and $t$ is the partial half-life of the superallowed 
$\beta$-decay branch. $\delta^{'}_R$, $\delta_{NS}$ and $\delta_C$ are the nucleus 
dependent part of the radiative correction, the nuclear structure dependent 
radiative correction, and the isospin-symmetry breaking correction,
respectively. $k$ is a constant ($k/(\hbar c)^6 = (8120.271\pm0.012) \times 
10^{-10} GeV^{-4}s$) and $\Delta_R^v$ is the transition independent part 
of the radiative corrections.

The corrected $Ft$ values allow to test the conserved vector current (CVC) 
hypothesis of the weak interaction. This hypothesis together
with the nine most precisely measured $0^+ \rightarrow 0^+$ transitions 
mentioned above yield an average $Ft$ value of 3072.2(8) s~\cite{towner02}.

The coupling constant $G_v$ together with the coupling constant for muon decay
allows to determine the $V_{ud}$ matrix element of the Cabbibo-Kobayashi-Maskawa
(CKM) quark mixing matrix, which in turn can be used to study the unitarity 
of the CKM matrix. This question has attracted much interest in recent years, 
as there were indications that the top row of the CKM matrix is not unitary
at the 2.2~$\sigma$ level~\cite{towner02,hardy90}. 
Recent new measurements~\cite{sher03} seem to indicate that the accepted value of 
$V_{us}$ might be too low. The new value of $V_{us}$, if confirmed, would restore
unitarity of the first row of the CKM matrix (0.9999(16) instead of 0.9968(14) before).
A deviation from unitarity 
would have far reaching consequences for the standard model of the
weak interaction and would point to physics beyond the currently accepted model.

Before the existence of physics beyond the standard model can be
advocated, the different inputs into the determination of the corrected $Ft$
value which leads to the calculation of the CKM matrix element should be 
carefully checked. It has turned out that the main uncertainty for the value of
the $V_{ud}$ matrix element comes from theoretical uncertainties linked to
the different correction factors. The correction which attracted mainly 
the attention is the isospin-symmetry breaking correction $\delta_C$, which 
is strongly nuclear-model dependent~\cite{towner02,ormand95,sagawa96}. 
Model calculations predict this 
correction to become increasingly important for heavier N=Z, odd-odd
nuclei. Measurements with these heavier nuclei should therefore allow to 
more reliably test this correction.

The aim of the work presented here is to determine with high precision the
half-life of $^{62}$Ga which is one experimental input to determine
the corrected $Ft$ value for this nucleus. This measurement is part of our
efforts which include also a measurement of the $\beta$-decay branching
ratios for $^{62}$Ga and improvements to the Canadian Penning Trap
(CPT) mass spectrometer to measure the mass of $^{62}$Ga and of its 
$\beta$-decay daughter $^{62}$Zn~\cite{savard03}.

The  half-life of $^{62}$Ga has been measured several times previously.
Alburger~\cite{alburger78} determined a value of 115.95(30)~ms, Chiba 
et al.~\cite{chiba78} obtained a less precise value of 116.4(15)~ms, 
Davids et al.~\cite{davids79} reported a value of 116.34(35)~ms, and 
Hyman et al.~\cite{hyman03} recently published a half-life of 115.84(25)~ms. 
These data yield a mean value 
of 116.00(17)~ms. This half-life does not yet reach the required precision of 
better than $10^{-3}$ in order to include $^{62}$Ga in the CVC test and 
the determination of the CKM matrix element. In the present paper, 
we report on a measurement which achieves the necessary precision.

\section{Experimental technique}

The experiment was performed at the GSI on-line mass separator. A $^{40}$Ca 
beam with an energy of 4.8 ~A~MeV and an average intensity of 50 particle-nA 
impinged on a niobium degrader foil of thickness 2.7~mg/cm$^2$ installed
in front of the silicon target of thickness 2.40~mg/cm$^2$. The niobium 
degrader decreased the energy of the primary beam to obtain the highest 
yield for the $^{28}$Si($^{40}$Ca,$\alpha$pn)$^{62}$Ga reaction. The beam 
intensity was controlled in regular time intervals by inserting a Faraday 
cup upstream of the degrader-target array. A Febiad-E2 source
was used to produce a low-energy $^{62}$Ga beam which was then mass analysed
by the GSI on-line mass separator and delivered to the measuring station.
The activity was accumulated on a moving tape device for 350~ms and then moved 
into the detection setup (transport time about 100~ms). After a delay of 10~ms, 
the half-life measurement was started. Measurement times of 1600~ms and 1800~ms
were used. During the measurement, the beam was deflected well ahead of the 
collection point. After about 2.2~s, a new cycle started with a new 
accumulation. 

The detection setup consisted of a 4$\pi$ gas detector used to detect
$\beta$ particles and a germanium detector for $\gamma$ rays.
The $\beta$ detector consists of two single-wire gas counters used in 
saturation regime and operated with P10 gas slightly above atmospheric 
pressure. The moving tape transporting the activity passed through
a slit between the two detectors. Thin aluminised mylar entrance 
windows allowed to detect low-energy $\beta$ particles.
The detectors were carefully tested before the experiment in order to establish 
their operation curves and to measure their efficiencies. With a $^{90}$Sr
source we determined an operation plateau between 1450~V and 1900~V and a 
detection efficiency of about 90\%. During the experiment, we used
different high voltage values within this plateau.
The detector volume was about 5 cm$^3$ for each detector.
The counting rate was found to be independent of the gas flow and a gas flow of 
about 0.15~$l/min$ was used throughout the experiment.

The germanium detector was mounted at a distance of about 4~cm from the source
point. It had a photopeak efficiency of 0.8\% at 1~MeV. Two $\gamma$-ray spectra
were accumulated in a multi-channel analyser in parallel during the run: 
i) one spectrum which required a $\beta$ coincidence from the gas detector
and ii) one spectrum where the germanium detector was running without any
coincidence.

The electronics chain for the $\beta$ detector was as follows: The output 
signal of the two detectors and a pulser signal were fed into one 
preamplifier (Canberra 2004). The detectors were biased by a Tennelec
HV power-supply (TC952A). The preamplifier output was connected to an ORTEC
timing-filter amplifier (TFA454) with integration and differentiation times of
100~ns and 20~ns, respectively. The TFA output was fed into a LeCroy octal 
discriminator (623B). In order to search for systematic effects in our data, we
modified the threshold during the experiment as described below. 
The output of the discriminator was split at the entrance of two LeCroy
Dual Gate Generators which defined the non-extendable dead-times of 3~$\mu$s 
and 5~$\mu$s, respectively. The output signal of the two gate generators 
were finally passed through a gate generator (GSI GG8000) and a level adapter
(GSI LA8010) to adjust the signal length and type for the clock (GSI TD1000)
which was started at the beginning of each cycle and which time-stamped each
event on a 2~ms/channel basis with 1000 channels. The time stamps were then
stored in a memory register (GSI MR2000). 
During the tape transport, the memory registers were read out, written to tape,
and erased for a new cycle.

The clock modules were carefully tested and selected before the experiment.
Measurements of their precision and stability have been carried out with 
respect to a Stanford Research Systems high precision pulse generator 
(DS345). Tests have been
performed at different counting rates (100~kHz to 900~kHz) and with different
times per channel (0.5~ms to 20~ms). The modules used had a timing 
precision of better than 2$\times$10$^{-6}$.

The background counting rate was checked before the experiment under 
experimental conditions. We measured a counting rate of 1.7~counts per second.
The average background determined in the half-life fits was about a factor 
of two higher (see below).

\section{Half-life measurements}

In this section, we will first discuss the results obtained for a reference
set of analysis parameters which yields the
final half-life value. In the second part, we will test how different analysis 
parameters may affect the half-life results. In the third part, we will 
investigate whether the data are subject to any systematic error. For this 
purpose, we varied different experimental parameters. Finally, the influence 
of different possible contaminants will be discussed.

\subsection{Analysis with reference parameters}

The data stored cycle-by-cycle have been analysed off-line with a program
package adapted from programs developped at Chalk River~\cite{koslowsky97}. 
This package includes a program to select cycles according to 
user defined criteria. The selection criteria are the number of counts in 
the spectrum for a given cycle and the $\chi^2$ value of a least-square 
fit of a user defined function (see below) to the data.
For a typical high-statistics run (e.g. run 10), about 10\% of the original data
has been rejected due to low statistics, whereas less than 1\% of the data
did not satisfy our fit quality criterium, i.e. either the fit did not converge
after 500 iterations or we obtained a too poor $\chi^2$ value. Therefore, if one 
excludes low-statistics cycles, at maximum 1\% of the cycles were rejected.
We do not expect that such a low rejection rate will bias our final result.

The fitting procedure is taken from a text book~\cite{bevington}.
The first condition rejects cycles where e.g. the primary beam was off or where
other technical problems reduced the counting rate significantly. The
second condition reject cycles with gas-detector sparks.
The reference conditions for this selection are a cutoff limit of 20 counts 
per cycle and a reduced $\chi^2$ value of 2 for the cycle fits. The fit 
was performed by assuming a one-component 
exponential and a long-lived fixed background, the latter being determined by 
an iterative procedure separately for each run. Variations of these parameters 
are discussed below.

For selected cycles, the program generates a simulated data set for which all
characteristics but the half-life are determined by the experimental cycle.
These simulated data are then subjected to a pre-defined dead-time
and stored cycle-by-cycle on disk. We used a half-life of 116.40~ms
in the simulations.

\begin{figure}[htb]
\begin{center}
\epsfig{file=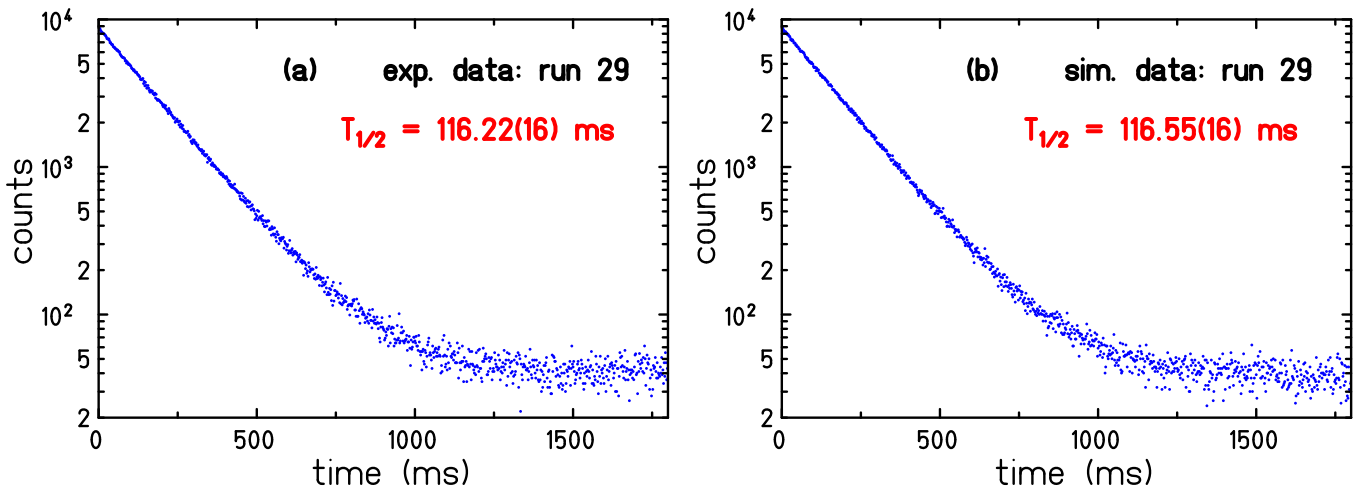,angle=-90,width=18cm}
\end{center}
\caption{(Color online) (a) Experimental decay spectrum for run 29. The spectrum has
         been corrected for dead-time. The half-life determined for this
         run is 116.22(16)~ms. (b) Simulated spectrum generated
         with the same characteristics as the experimental spectrum, 
         but an input half-life of 116.40~ms. After dead-time correction, 
         the half-life determined is 116.55(16)~ms.
}
\label{fig:spec}
\end{figure}

The second step in the analysis procedure is a correction cycle-by-cycle
of the dead-time. This is achieved by a correction of each channel with a 
factor of $f_{dt} = 1 / (1+t_{dt}R)$, where $t_{dt}$ is the non-extendable
dead-time per event (3$\mu$s or 5$\mu$s) and $R$ is the counting rate (1/s)
determined from the channel content. After dead-time correction, all selected 
cycles are summed to yield the decay-time spectrum of a given run. 39 different
runs with run times as long as 4~h at maximum and varying experimental
conditions (high voltage of the gas detector, CFD threshold, cycle time, 
transport tape) were accumulated. In total, about 38$\times$10$^6$ $^{62}$Ga
decays have been detected.

In figure~\ref{fig:spec}a, we show the experimental spectrum of one run
corrected for dead-time. The fit of this particular run which lasted about 4h
yielded a half-life of 116.22(16)~ms. Part~b of the figure shows
the simulated spectrum. Here a half-life of 116.55(16)~ms resulted.

\begin{figure}[htb]
\begin{center}
\epsfig{file=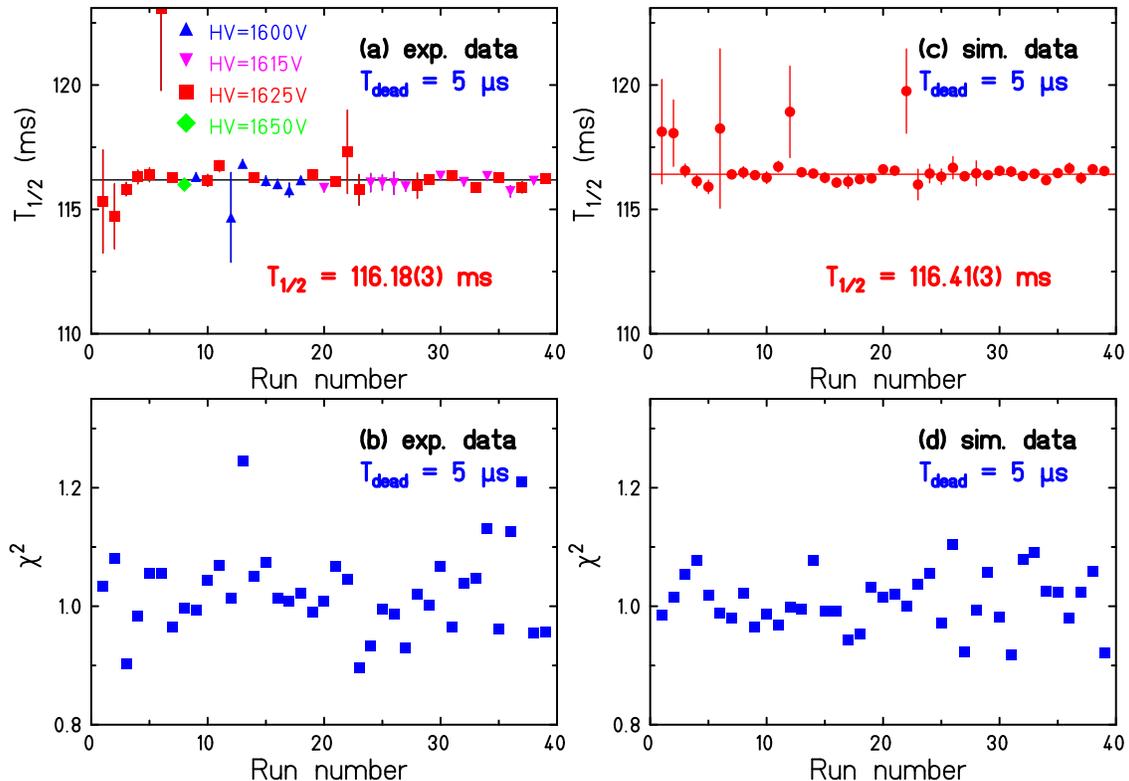,angle=-90,width=15cm}
\end{center}
\caption{(Color online) (a) Experimental half-lives as determined for the different runs.
          The results shown originate from the data set with a preset dead-time 
          of 5$\mu$s. The average results in a half-life value of 116.18(3)~ms. 
          The reduced  $\chi^2$ value for the averaging procedure is 1.6.
         (b) Reduced $\chi^2$ values for the fits of the 
          experimental spectra of the individual runs.
         (c) and (d) show the same information as (a) and (b), however, for
          the simulated decay distributions. Here a half-life of 116.41(3)~ms
          is obtained, in agreement with the input half-life of 116.40~ms.
          The averaging procedure yields here a reduced $\chi^2$ of better than
          unity.
}
\label{fig:exp}
\end{figure}

Similar results were obtained for the other runs.
Figure~\ref{fig:exp}a shows the half-life determined for each of the 39 runs
for the data set with a dead-time of 5~$\mu$s. The fits of the individual 
runs were performed with a decay component 
for $^{62}$Ga and a constant background. The error-weighted average 
of the half-life is 116.18(3)~ms. These data have been analysed with the 
reference conditions defined above. The reduced $\chi^2$ value of the 
averaging fit is 1.60. 

Figure~\ref{fig:exp}b shows the reduced $\chi^2$ value of the fits of the 
individual runs. In general, the values are below 1.1. Two runs are at a 
reduced $\chi^2$ value of about 1.13 and two more above 1.2. These high 
$\chi^2$ values indicate a rather low probability (below 0.1\%) 
for the fitted distribution being the true distribution. 
We analysed these runs very carefully by e.g. cutting them in
up to four "sub-runs" to see whether there is any indication
of increased detector sparking or any other experimental problem.
As we did not find any difference 
between these runs and those with lower $\chi^2$ values, we assume that these
high values are a result of the tails of the probability distribution of
$\chi^2$. Removal of these runs would not have changed the final result by 
any means.

In figures~\ref{fig:exp}c,d, we plot the same information for the simulated 
data. The average value for the half-life is 116.41(3)~ms in agreement with the 
input value of 116.40~ms. Most of the reduced $\chi^2$ values are close to 
unity, but there are again runs with higher values from the tails of the 
$\chi^2$ probability distribution. The fact that the half-life result for the 
simulated data corresponds to the input half-life shows that the dead-time 
correction is done correctly. An omission of the dead-time correction for the 
5$\mu$s data would yield a half-life that is longer by about 0.15~ms. 

\begin{figure}[htb]
\begin{center}
\epsfig{file=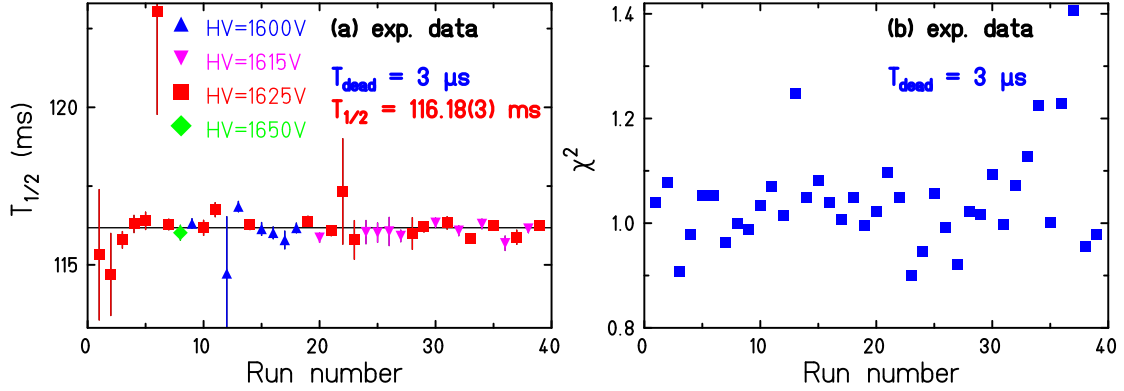,angle=-90,width=15cm}
\end{center}
\caption{(Color online) (a) Experimental half-lives as determined from the data set registered 
             with a fixed dead-time of 3~$\mu$s. The error weigthed average
             value is 116.18(3)~ms with a reduced $\chi^2$ of 1.65.
             (b) Reduced $\chi^2$ values for the fits of the 
             different runs.
}
\label{fig:3mus}
\end{figure}

The data with a fixed dead-time of 3$\mu$s were analysed in a similar way, 
including again a comparison with simulations. 
The results are in perfect agreement 
with the 5$\mu$s data set. In figure~\ref{fig:3mus}, we show the half-life 
values for the individual runs as deduced from this data set. The average value 
is 116.18(3)~ms. The reduced $\chi^2$ distribution is very similar to the one 
of the first data set. The fact that we obtain the same results for
the two data sets with different dead-time values indicates again that the
dead-time correction in the data analysis is handled properly.

\begin{figure}[hbb]
\begin{center}
\epsfig{file=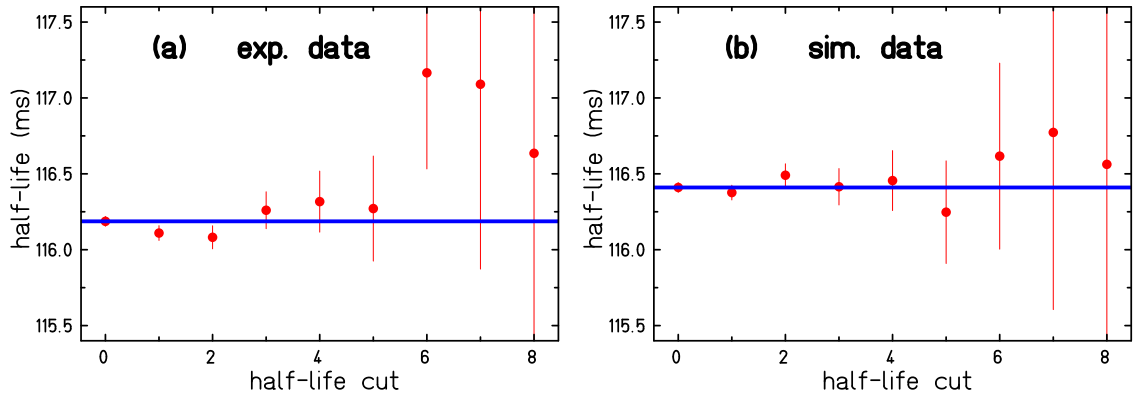,angle=-90,width=15cm}
\end{center}
\caption{(Color online) (a) Experimental half-lives as determined for cuts in the time
             distribution. The data point shown at an x value of zero
             is the half-life obtained when the whole spectrum is fitted.
             The data point at an x value of 1 originates from a fit where
             the first 116~ms, i.e. the first half-life of $^{62}$Ga, of the
             spectrum were omitted etc. The solid line corresponds to the 
             half-life value without cut.
             (b) Same cuts performed on the simulated data.
             The data used for this analysis include all runs with a 5$\mu$s 
             dead-time.
}
\label{fig:cuts}
\end{figure}

Another way to track down possible problems with the dead-time correction
is to fit the experimental and simulated data for different parts of the 
spectra. This uses the fact that the data at the beginning of the cycle are, 
due to the higher counting rate, more strongly influenced by the dead-time than
the data later in the cycle. In figure~\ref{fig:cuts}, we present this analysis
of the data for the two data sets, experiment and simulations.
The half-life shows indeed very little variation as a function of the
starting point of the fit and the half-lives obtained vary around the value
obtained when the whole spectrum is fitted.

\subsection{Influence of analysis parameters}

The selection of cycles can be performed with different parameter values.
This selection may influence the final result considerably. We therefore
tested the different selection criteria extensively.

The most important selection parameter is the $\chi^2$ value of the fit of the
individual cycles. We performed tests with reduced $\chi^2$ limits of 1.5, 2, 5, 
20, 2000, and 20000. The reduced $\chi^2$ value of the cycle fits has to be equal to or 
smaller than the chosen $\chi^2$ value to be taken into account for further 
analysis. No significant change of the half-life obtained was observed for
values of 2 and higher. However, a too tight limit on $\chi^2$ yields longer 
half-lives. A reduced $\chi^2$ limit of 1.5 in the cycle selection yields
a half-life about 0.2\% longer than for higher values. 

The use of a reduced $\chi^2$ value higher than two does not alter the 
resulting final half-life, however, the $\chi^2$ value of the individual 
runs increases significantly with increasing $\chi^2$ limit, as less and 
less detector sparks are eliminated. We finally chose a reduced 
$\chi^2$ value of 2 to efficiently remove cycles with detector sparks.

An additional selection parameter is the number of counts for a cycle.
We varied this parameter but could not find any systematic
behaviour as a function of its value. The final analysis was performed with
a value of 20. This removes basically all cycles with low activity.

As mentioned above, the cycle selection was performed by  fitting with only 
one decay component and a fixed constant background. The fixed constant
background option was chosen to avoid negative background fit values that may
occur due to the very low statistics per cycle. The fixed background was
determined iteratively for each run individually. This procedure converged 
after only two or three iterations. We performed also analyses with 
background values different from those determined by the iterative procedure.
It turned out that slightly different half-lives were obtained only when
reduced $\chi^2$ values smaller than two were used.

To test the fitting procedure, we performed also fits of our data with the 
MINUIT~\cite{minuit} fit package. These fits performed for individual runs
as well as for a sum spectrum of all runs yield, after dead-time correction, 
precisely the same results as the least-square procedure from a text 
book~\cite{bevington} used in the main analysis.

\subsection{Search for systematic effects due to experimental conditions}

During the experiment, certain experimental parameters were changed to
study their influence on the half-life determined. For this purpose, we used
two different values for the CFD threshold which triggered the decay events.
Runs up to run  number 17 were performed with a threshold value of -0.3~V.
For the 5$\mu$s data, these runs give an average half-life of 116.26(6)~ms.
The runs from number 18 on used a slightly different value of -0.35~V.
Here the average half-life is 116.14(4)~ms. For the 3$\mu$s data, similar 
values were obtained. These values agree with each other at the 1.4~$\sigma$ 
level.

Another parameter changed periodically during the data taking is the
detector high-voltage. The values used varied between 1600~V and 1650~V.
In figure~\ref{fig:exp}a,b, the runs with different high-voltage values 
are shown by different symbols. No systematic effect was observed.
The average half-lives determined are 116.28(7)~ms, 116.10(6)~ms, 116.21(5)~ms,
and 116.00(24)~ms for the high-voltage values of 1600~V, 1615~V, 1625~V, 
and 1650~V, respectively. The results for the 3$\mu$s data set is in agreement 
with the 5$\mu$s data.

Runs 1-14 were performed with a decay time of 1600~ms, whereas a decay time
of 1800~ms was used for the rest of the runs. This change did not induce
any observable systematic effect. Finally the transport tape was changed
after run 14 to ensure that long-lived activities do not alter the results.
No effect was observed.

Within the tests performed our data are not subject to any experimental
bias. We therefore do not correct our data for systematic errors
due to experimental parameters.

\subsection{Search for contaminants}

Contaminants which are produced in the same reaction and transported
in the same way as $^{62}$Ga may alter the observed half-life significantly.
In general, their half-life is different from the half-life of $^{62}$Ga
and their influence has to be studied carefully.

In an isobaric chain separated by a mass separator, the half-lives
change significantly from one isotope to the other. As the less exotic
isotopes have normally longer half-lives and are usually produced with 
much higher cross sections, an isobaric contamination not corrected for 
properly yields longer half-lives.

Our data are possibly subject to different kinds of contamination:

\begin{itemize}
\item
As the mass separation is not perfect, isotopes with masses A=61 or A=63 might 
contaminate our data. However, at the GSI on-line separator neighboring masses
are suppressed by a factor of the order of $10^4$ which does not take 
differences in production cross-section and half-life into account. This high 
selectivity is confirmed by the fact that no $\gamma$ rays from mass 
61 or 63 isotopes were observed in the $\beta$-particle gated
$\gamma$ spectra which yields a suppression factor well above 10$^4$.

\item
$^{62}$Zn ($T_{1/2}$=9.2h) is produced i) directly in the ion source of the mass separator and 
transported to the collection point and ii) by the decay of $^{62}$Ga. However, 
due to its low $Q_{EC}$ value of 1627~keV, $^{62}$Zn decays mainly by 
electron capture. As the half-life of $^{62}$Ga was determined
by measuring the time characteristics of $\beta$ particles, $^{62}$Zn is not 
expected to alter the half-life value of $^{62}$Ga. In addition, its half-life
is very long compared to that of $^{62}$Ga so that its contribution 
is in the worst case expected to yield a constant background. These findings
are confirmed by the absense of $\gamma$ rays from the decay of $^{62}$Zn
in the $\beta$-gated $\gamma$ spectrum.

\item
$^{62}$Cu ($T_{1/2}$=9.7 min) is certainly the most important contaminant because
its half-life is short enough to be a concern. It is either produced directly in 
the ion source of the mass separator and implanted as a mass-separated beam 
or produced in the decay of the $^{62}$Zn activity collected on the tape.
The latter contribution can safely be discarded, as it is governed by the  
$^{62}$Zn half-life which is very long. 

To determine the influence of $^{62}$Cu produced in the mass separator source 
and collected together with $^{62}$Ga, we attempted to fit the time 
distributions for the individual runs as well as for the sum of all runs
by adding a second exponential with the time constant of $^{62}$Cu to the
fit function. However, the fit reduces this contribution to a value in agreement
with zero. If we impose a $^{62}$Cu contribution by fixing the amount of
$^{62}$Cu, the fit reduces the background, but keeps the $^{62}$Ga 
half-life constant. In fact the half-life changes by 0.01~ms, if we compare 
a fit without any $^{62}$Cu contribution and a fit with the highest possible
contribution of $^{62}$Cu which is defined by
the fact that no constant background is left over in the fit. As the 
constant background cannot become negative, the $^{62}$Ga half-life
becomes shorter if the $^{62}$Cu contribution is still increased. In the same 
way, the reduced $\chi^2$ value of the fit stays unity until 
the background is zero and the
$^{62}$Ga half-life becomes shorter. This finding means that a possible 
$^{62}$Cu contribution is comparable to a constant background and we can
not distinguish both, i.e. a possible $^{62}$Cu contamination does not alter
the $^{62}$Ga half-life. These results are shown graphically in 
figure~\ref{fig:cu62}.

\begin{figure}[htb]
\begin{center}
\epsfig{file=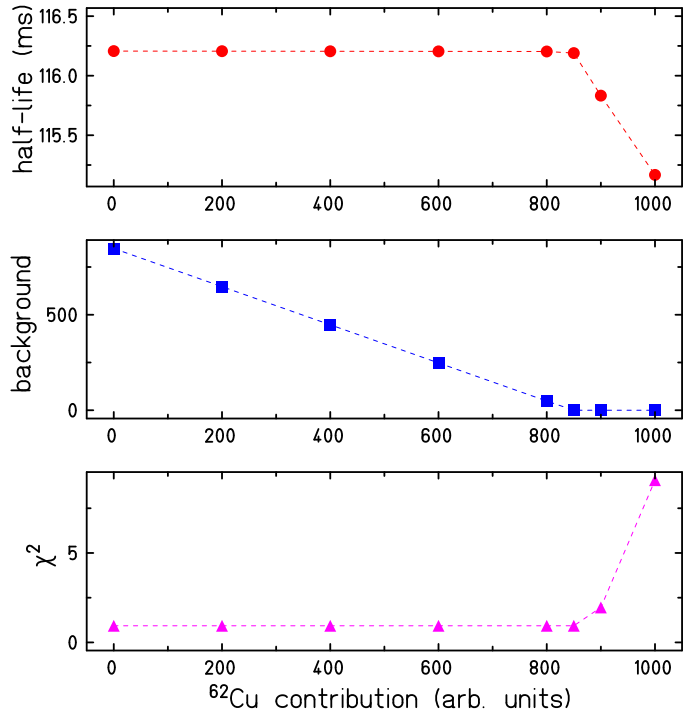,angle=0,width=9cm}
\end{center}
\caption{(Color online) Part (a) shows the variation of the $^{62}$Ga half-life as a function
         of the amount of contamination by $^{62}$Cu imposed in the fit.
         Part (b) indicates how the constant background decreases as the 
         $^{62}$Cu contribution increases, whereas part (c) demonstrates
         that the fit quality is not altered as long as the constant background 
         can be reduced by the fit. The reduced $\chi^2$ value and the half-life
         of $^{62}$Ga change only, once the constant background has vanished
         and the $^{62}$Cu contribution is still increased. The whole data set
         with a dead-time of 5$\mu$s is used for this analysis.
}
\label{fig:cu62}
\end{figure}

We searched also for a $^{62}$Cu contamination in the $\beta$-gated $\gamma$
spectrum. However, none of the known $\gamma$ rays of $^{62}$Cu could be
identified which supports the conclusion that $^{62}$Cu does not play an 
important role in the present experiment.

\item
$^{62}$Ge ($T_{1/2}$=129$\pm$35~ms~\cite{lopez02}) can also be produced by the 
$^{28}$Si($^{40}$Ca,$\alpha$2n) reaction used in the
present experiment. The production of this more exotic nucleus, however, 
is expected to be significantly smaller. Although fusion-evaporation codes
do not very reliably predict very small cross sections, it is worth 
mentioning that the HIVAP code~\cite{hivap} predicts $^{62}$Ge with yields 
which are a factor of $10^{-5}$ smaller than those for $^{62}$Ga. In addition, 
the release efficiency of the Febiad-E2 ion source for germanium isotopes is 
known to be much smaller than that for gallium isotopes which reduces 
the probability of an important $^{62}$Ge 
contamination even more. As the half-life of $^{62}$Ge 
is rather close to the one of $^{62}$Ga, large amounts of $^{62}$Ge would be 
necessary to alter the half-life determined for $^{62}$Ga. We therefore
conclude that the influence of $^{62}$Ge is negligible.
\end{itemize}

\subsection{Final experimental result}

The evaluation of possible systematic errors from contaminants or other sources
shows that no such corrections to the half-life are necessary. 
Therefore, our experimental result with its statistical 
error is 116.18(3)~ms. However, when averaging the different runs we obtain
a reduced $\chi^2$ value of 1.6 which indicates that there might be some hidden 
inconsistencies between the different runs. A possible explanation could be
that our cycle selection procedure does not remove all detector sparks. 
The probability that the fit function (a constant in our case) is the correct 
function to describe the data is only 1\% with a reduced $\chi^2$ value of 
1.6~\cite{bevington}.

We therefore apply a procedure proposed by the Particle Data 
Group~\cite{pdg} and multiply the error bars of the individual runs 
by the square-root of the reduced $\chi^2$ value obtained from the averaging
procedure. The subsequent averaging gives the final experimental
value of 116.19(4)~ms. 

\section{Discussion}

Our experimental result of 116.19(4)~ms has to be compared with other
half-life measurements from the literature. In table~\ref{tab:half}, we list
all published half-life measurements for $^{62}$Ga. Our result is in excellent 
agreement with all measurements, but the most recent publication by Hyman et 
al.~\cite{hyman03}. In fact, the discrepancy of the present result with the 
one of Hyman et al. is at the 1.2~$\sigma$ level. When averaging all 
experimental values, we obtain an error weighted average value of 116.18(4)~ms
with a reduced $\chi^2$ value of 0.68.

Three measurements for the $\beta$-decay branching ratio of $^{62}$Ga have been 
published recently~\cite{hyman03,blank02ga62,doering02}. All measurements yield
consistent results of 0.120(21)\%, 0.12(3)\%, and 0.106(17)\% for the population
and decay of the first excited state in $^{62}$Zn. Hyman et al.\cite{hyman03} 
used these results and shell-model calculations to determine the superallowed 
branching ratio to be 99.85$^{+0.05}_{-0.15}$\%. By using this result 
together with the measured Q value~\cite{davids79} of 9171(26)~keV and the 
half-life of 116.18(4)~ms, we obtain an $ft$ value of $ft$~=(3056$\pm$47)s 
for $^{62}$Ga. When taking into account the correction factors as calculated 
by Towner and Hardy~\cite{towner02}, we finally get a corrected $Ft$ value 
of $Ft$~= (3056$\pm$47)s (the $Ft$ value numerically does not change compared
to the $ft$ value, because the correction factors $\delta^{'}_R$ and 
$\delta_c - \delta_{NS}$ are similar in size). This value is in agreement 
with the average value from the nine precisely measured transitions between
$^{10}$C and $^{54}$Co of $Ft$~=3072.2(8)~s~\cite{towner02}.

The large uncertainty of our corrected $Ft$ value comes almost entirely from
the error associated with the poorly known $\beta$-decay Q value. 
The correction factors can therefore not yet be tested with the experimental
precision reached up to now. A much more precise value for the Q value 
is needed for this purpose. However, assuming that CVC holds and
the correction factors are correct, we can determine the Q value
for the decay of $^{62}$Ga. We obtain a value of 9180(4)~keV which is consistent
with the value determined by Hyman et al.~\cite{hyman03} of 9183(6)~keV.

\section{Conclusions}

We have performed a high-precision measurement of the half-life of $^{62}$Ga.
The half-life was determined by detecting the $\beta$ particles from
a $^{62}$Ga source produced at the GSI on-line mass separator. The result 
of T$_{1/2}$~= 116.19(4)~ms obtained in this work is in agreement with 
older half-life values from the literature. The present result is more than
a factor of four more precise than any previous result. We determine
an error-weighted mean value of 116.18(4)~ms for the half-life of $^{62}$Ga.

This half-life value is now precise enough to contribute to a stringent 
test of CVC above Z=27 as soon as a more precise $\beta$-decay Q value
and a refined value of the $\beta$-decay branching ratios are known.
$^{62}$Ga can then also be used to test the correction factors
calculated to determine the nucleus independent $Ft$ value.
Work on these issues is under way.

\section*{Acknowledgements}

We acknowledge the help of K. Burkard and W. H\"uller during the data 
taking at GSI and the work of the GSI accelerator crew for delivering a 
high-intensity, high-quality beam. This work was supported in part by the US 
Department of Energy under Grant No. W-31-109-ENG-38, by the R\'egion 
Aquitaine, and by the European Union under contract HPRI-CT-1999-50017.

\begin{table}[hht]
\begin{center}
\begin{tabular}{|ccccc|c|}
\hline \rule{0pt}{1.3em}
present work  & Alburger~\cite{alburger78} & Chiba et al.~\cite{chiba78} &
Davids et al.~\cite{davids79} & Hyman et al.~\cite{hyman03}~~~~ & average \\
[0.5em]
\hline \rule{0pt}{1.3em}
116.19(4) ms & 115.95(30)~ms & 116.4(15)~ms & 116.34(35)~ms & 115.84(25)~ms &
116.18(4) ms \\
[0.5em]
\hline
\end{tabular}
\caption{Results of half-life measurements for $^{62}$Ga. The last column shows 
         the error weighted average value of all measurements.}
\label{tab:half}
\end{center}
\end{table}

\end{document}